\documentclass[superscriptaddress,twocolumn,prl,showpacs]{revtex4}
\usepackage{amsmath}
\usepackage{amssymb}
\usepackage{amsfonts}
\usepackage{euscript}
\usepackage{graphicx}

\newcommand{\be}[1]{\begin{equation}\label{#1}}
\newcommand{\ee}{\end{equation}}
\newcommand{\ba}[1]{\begin{eqnarray}\label{#1}}
\newcommand{\ea}{\end{eqnarray}}

\newcommand{\etal}{{\it et al.}}
\newcommand{\dd}{\dagger}

\def\II{\hbox{$1\hskip -1.2pt\vrule depth 0pt height 1.6ex width 0.7pt\vrule depth 0pt height 0.3pt width 0.12em$}}
\def\RR{\mathbb{R}}

\def\G{\Gamma}

\newcommand{\cD}{\mathcal{D}}
\newcommand{\cE}{\mathcal{E}}

\newcommand{\cH}{\mathcal{H}}

\newcommand{\cP}{\mathcal{P}}
\newcommand{\cT}{\mathcal{T}}

\begin{document}

\title{$\cP\cT$ symmetry and spontaneous symmetry breaking in a microwave billiard}

\author{S.~Bittner}
\affiliation{Institut f{\"u}r Kernphysik, Technische Universit{\"a}t
Darmstadt, D-64289 Darmstadt, Germany}

\author{B.~Dietz}
\email{dietz@ikp.tu-darmstadt.de}
\affiliation{Institut f{\"u}r Kernphysik, Technische Universit{\"a}t
Darmstadt, D-64289 Darmstadt, Germany}

\author{U.~G\"unther}
\affiliation{Helmholtz-Zentrum
Dresden-Rossendorf, Postfach 510119, D-01314 Dresden, Germany}

\author{H.~L.~Harney}
\affiliation{Max-Planck-Institut f{\"u}r Kernphysik, D-69029 Heidelberg,
Germany}

\author{M.~Miski-Oglu}
\affiliation{Institut f{\"u}r Kernphysik, Technische Universit{\"a}t
Darmstadt, D-64289 Darmstadt, Germany}

\author{A.~Richter}
\email{richter@ikp.tu-darmstadt.de}
\affiliation{Institut f{\"u}r Kernphysik, Technische Universit{\"a}t
Darmstadt, D-64289 Darmstadt, Germany}
\affiliation{$\rm ECT^*$, Villa Tambosi, I-38123 Villazzano (Trento), Italy}

\author{F.~Sch{\"a}fer}
\affiliation{Institut f{\"u}r Kernphysik, Technische Universit{\"a}t
Darmstadt, D-64289 Darmstadt, Germany}
\affiliation{LENS, University of Florence, I-50019 Sesto-Fiorentino (Firenze),
Italy}

\date{\today}

\begin{abstract}
We demonstrate the presence of parity-time ($\cP\cT$) symmetry for the non-Hermitian two-state Hamiltonian of a dissipative microwave billiard in the vicinity of an exceptional point (EP). The shape of the billiard depends on two parameters. The Hamiltonian is determined from the measured resonance spectrum on a fine grid in the parameter plane. After applying a purely imaginary diagonal shift to the Hamiltonian, its eigenvalues are either real or complex conjugate on a curve, which passes through the EP. An appropriate basis choice reveals its $\cP\cT$ symmetry. Spontaneous symmetry breaking occurs at the EP.
\end{abstract}
\pacs{02.10.Yn, 05.45.Mt, 11.30.Er} \maketitle

{\em Introduction.}\quad Flat microwave cavities, so called "microwave billiards" are analogues of quantum billiards~\cite{QB}. They are an experimental test ground for the properties of the eigenvalues and eigenfunctions of quantum systems~\cite{Richter}. In the present Letter we demonstrate theoretically and experimentally that they can also be used to study dissipative quantum systems which have a parity-time ($\cP\cT$) symmetry, that is, are invariant under the simultaneous action of a parity operator ($\hat P$) and a time reversal operator ($\hat T$) after a suitable width-offset.

Strong interest in $\cP\cT$-symmetric quantum systems was initiated in 1998 by Ref.~\cite{cmb-pt-prl-1998} demonstrating that a non-Hermitian Hamiltonian $\cH$ has real eigenvalues provided it respects $\cP\cT$ symmetry, i.e. $[\hat P\hat T,\cH]=0$ and has eigenvectors that are also $\cP\cT$ symmetric. In \cite{bender-berry-mandilara} this statement was generalized to systems invariant under an antilinear operator $\hat T^\prime=\hat U\hat T$, where $\hat U$ is unitary. The observations in Refs.~\cite{cmb-pt-prl-1998,bender-berry-mandilara} led to a reconsideration of the necessity of the Hermiticity axiom for quantum observables~\cite{cmb-rev}. Depending on external parameters the $\cP\cT$ symmetry of the eigenvectors may be spontaneously broken, i.e. they cease to be eigenvectors of $\hat P\hat T$, although $\cH$ still commutes with $\hat P\hat T$~\cite{cmb-pt-prl-1998,cmb-rev}. As a result, the eigenvalues of $\cH$ are no longer real, but rather become complex conjugate pairs. This phase transition occurs at an exceptional point (EP)~\cite{kato} defined as the coalescence of at least two eigenvalues and the corresponding eigenvectors. 

Close to an EP, non-Hermitian but $\cP\cT$-symmetric Hamiltonians are capable to speed-up quantum evolution processes~\cite{PT-brach}. During the last years optical wave-guide systems with fine-tuned ($\cP\cT$-symmetrically balanced) gain and loss regions, i.e.~with active (laser pumped) and absorptive components, have been investigated theoretically~\cite{PT_Theorie}. This led to experimental setups which consist of one wave-guide with absorption and another one which is either optically pumped (active $\cP\cT$ symmetry)~\cite{gain-loss5} or lossless (passive $\cP\cT$ symmetry)~\cite{guo-passive}. In optical systems the $\cP\cT$ phase transition has been studied theoretically~\cite{gain-loss3} as well as experimentally~\cite{gain-loss5,guo-passive}. 
        \begin{figure}[!t]
        \centering
        \includegraphics[width=5cm]{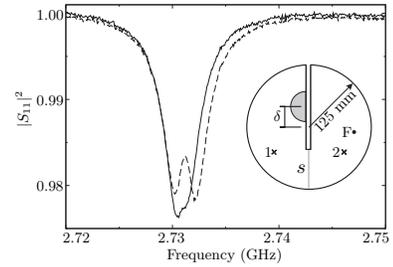}
        \caption{ Two typical spectra 
$\vert S_{11}(f)\vert^2$ for $B=38$ mT and the parameter settings $(s,\delta)=(1.66,41.79)$~mm (solid line), and $(s,\delta)=(1.99,41.76)$~mm (dashed line), respectively, in the vicinity of the EP, which is located at $(s_{\rm EP},\delta_{\rm EP})=(1.72\pm 0.01,41.78\pm 0.01)$~mm. Inset: top view (to scale) of the microwave billiard. The setup is described in the text.}
        \label{fig:1}
\end{figure}
Recently it was observed in a non-Hermitian system with active $\cP\cT$ symmetry consisting of one amplifying and one attenuating LRC circuit~\cite{Kottos_2011}. 
Further theoretical studies of $\cP\cT$ symmetry based effects concern spectral singularities~\cite{most-prl-spec-sing-prl1}, lasers at threshold~\cite{most-prl-spec-sing-pra2,stone-laser-absorber-prl}, coherent perfect absorbers~\cite{stone-laser-absorber-prl}, unidirectional invisibility induced by $\cP\cT$-symmetric periodic structures~\cite{dim-kott} as well as Bloch oscillations in $\cP\cT$-symmetric lattice structures~\cite{longhi-bloch-osc-prl} and optical tachyons in $\cP\cT$-symmetric optical graphene-like structures~\cite{segev-pt-graphene-tach}. 
The $S$-matrix formalism for $\cP\cT$-symmetric systems was analyzed recently~\cite{stone-laser-absorber-prl,schomerus-prl}.

{\em Experiment.}\quad We will show that configurations with passive $\cP\cT$ symmetry, including a $\cP\cT$ phase transition, are observable in a dissipative microwave billiard, whose shape depends on two parameters. The experiments were performed in the vicinity of an EP in the parameter plane. The setup was similar to that used in~\cite{prl_2001}. There the first experimental evidence of EPs and of geometric phases was provided. Recently, these experiments have been extended to systems with induced $\cT$ violation~\cite{prl_2011}. The obtained data are used for the present work. The experimental setup (see the inset of Fig.~\ref{fig:1}), described in \cite{prl_2011}, consisted of a flat, 5~mm high microwave resonator of circular shape with 125~mm radius. It was divided into two approximately equal parts by a 10~mm thick copper bar, with an opening of 80~mm.
The coupling between the electric field modes in each part via the opening was varied with a copper gate that had a tilted bottom. It was inserted into the resonator through a slit in its top and could be moved up and down. The bottom plate had a notch to enable a complete closing of the gate. The lifting $s$ defines one parameter, where $0\ {\rm (no\ coupling)} \leq s \leq 9~{\rm mm\ (maximal\ coupling)}$. The second parameter is provided by the displacement $\delta$ (with respect to the cavity center) of a 5~mm high semicircular Teflon piece of radius 30~mm inserted into the left part of the cavity. The parameter plane $(s,\delta )$ was scanned with the help of two micrometer stepper motors, which moved the gate and the Teflon piece in steps of $\Delta s = \Delta \delta = 0.01~{\rm mm}$. Furthermore, a 5~mm high cylindrical ferrite of radius 2~mm denoted by F in Fig.~\ref{fig:1} was inserted into the right part of the cavity. To induce $\cT$ violation the ferrite was magnetized with an external magnetic field $B$ of strength $0\leq B \leq 90~{\rm mT}$, applied perpendicularly to the billiard plane~\cite{prl_2007,prl_2011}. Two pointlike wire antennas $1$ and $2$ reached into the cavity, one into its left, the other into its right part. A vectorial network analyzer Agilent PNA 5230A emitted microwave power into the resonator via one antenna $a$ and received an output signal either at the same or at the other antenna $b$. It determined the amplitude and phase of the output signal relative to the input signal, thus yielding the four complex elements $S_{ba}(f)$, $\{a,b\}\in\{1,2\}$ of the scattering matrix $S(f)$. They were measured for each setting of the parameters $(s,\delta)$ as a function of the excitation frequency $f$ in steps of $\Delta f =10~{\rm kHz}$. The frequency range of 40 MHz was determined by the spread of the resonance doublet under consideration. In that range the electric field vector is perpendicular to the top and bottom plates. Therefore, the Helmholtz equation is mathematically identical to the Schr\"odinger equation of the quantum billiard~\cite{QB,Richter}. Thus, the results of this Letter also apply to the associated quantum system. 
Figure~\ref{fig:1} shows two reflection spectra measured for $B=38$~mT, one for $s<s_{\rm EP}$ (solid line) and one for $s>s_{\rm EP}$ (dashed line). For the former we observe a close encounter of the resonance positions, for the latter one of the widths. In fact, we observe this feature of the resonance spectra along a curve in the parameter space which crosses the EP independently of the choice of the external magnetic field. Exactly at the EP the resonance shape is that of a second order pole in addition to the first order poles~\cite{Hernandez}. As demonstrated in the following, along this curve the effective Hamiltonian has the form of a $\cP\cT$-symmetric Hamiltonian after a suitable basis transformation. 

Neighboring resonances are situated about $250~{\rm MHz}$ away from the doublet under consideration. Consequently, the effective Hamiltonian $H^{\rm eff}$ is two-dimensional. It is determined for every setting of $s,\delta ,{B}$ by fitting an analytic expression for the $S$-matrix~\cite{prl_2007,mahaux-weidenmueller} originally derived in the context of nuclear reaction theory and extended to microwave resonators in~\cite{Albeverio}, 
\begin{equation}
S_{ab}(f)=\delta_{ab}-2\pi i\sum_{\mu,\nu=1}^2W^\star_{a\mu}\left[(f\II-H^{\rm eff})^{-1}\right]_{\mu\nu}W_{b\nu},\label{eq:1}\\ 
\end{equation}
to the measured one. The effective Hamiltonian
\begin{equation}
H^{\rm eff}_{\mu\nu}=H_{\mu\nu}-i\pi\sum_c W_{c\mu}W^\star_{c\nu}\label{eq:1a}
\end{equation}
is obtained by evaluating the integrals entering Eq.~(4.2.20b) of Ref.~\cite{mahaux-weidenmueller} and Eq.~(4) of \cite{prl_2007}. The matrix $\II$ is the unit matrix and $H$ is the Hermitian two-state Hamiltonian of the closed resonator or, equivalently, the quantum billiard. It includes the coupling of the ferromagnetic resonance to the rf magnetic field in case of a nonzero external magnetic field $B$ and thus takes account of the violation of $\cT$ invariance. The matrix elements $W_{a\mu}, \, W_{b\mu}$ couple the resonator modes $\mu =1,2$ to the antenna states $\{a,b\}\in\{1,2\}$ and the sum over $c$ includes the fictitious channels, which describe dissipation in the walls of the resonator and the ferrite~\cite{prl_2009,prl_stoeckmann03}. These matrix elements are real and frequency independent in the considered range. Thus, for a vanishing external magnetic field $H^{\rm eff}$ is given by Eq.~(\ref{eq:1a}) with real matrix elements $W_{a\mu}$ and a time reversal invariant $H$, $[\hat T,H]=0$. Here, $\hat T$ is the antilinear operator of complex conjugation, whence $H$ is real symmetric. As a consequence, the scattering process is also time-reversal invariant, $\hat TS\hat T=S^\dd$~\cite{henley}, and the $S$ matrix is symmetric, $S_{ab}= S_{ba}$. $\cT$ violation is induced with a nonvanishing $B\ne 0$. In this case, the coupling of the resonator states to the ferromagnetic resonance is complex. Consequently, the evaluation of the integrals entering Eq. (4) of Ref.~\cite{prl_2007} yields a complex and antisymmetric contribution, leading to a complex Hermitian Hamiltonian $H$ in Eq.~(\ref{eq:1a}), which is not invariant with respect to $\hat T$. Lack of reciprocity, i.e. $S_{ab}\ne S_{ba}$, in the measured spectra is the signature for $\cT$ violation~\cite{prl_2007,prl_2011}. In the $\cT$-invariant case $H^{\rm eff}$ is complex symmetric, otherwise it is nonsymmetric. 
To simplify notation we express it most generally in terms of the unit matrix $\II$ and the Pauli matrices $\vec\sigma=(\sigma_x,\, \sigma_y,\, \sigma_z)$ as (see Eq.~(1) of Ref.~\cite{prl_2011}),
\begin{equation}
H^{\rm eff}=\frac{e_1+e_2}{2}\II +\vec\sigma\cdot\vec h,\,\,
\vec h=\left(H^S_{12},H^A_{12},(e_1-e_2)/2\right).
                   \label{eq:2}
\end{equation}
All entries of $H^{\rm eff}$ are complex, and depend on $(s,\delta)$, but not on $f$ in the considered frequency range. Fitting the $S$ matrix Eq.~(\ref{eq:1}) to the measured one determines $H^{\rm eff}$ and $W_{1\mu},\, W_{2\mu}$ up to common real orthogonal basis transformations. As in \cite{prl_2011} we choose the basis such that the ratio of the off-diagonal elements of $H^{\rm eff}$ equals
\begin{equation}
\frac{H^S_{12}+iH^A_{12}}{H^S_{12}-iH^A_{12}}=e^{2i\tau},\, \tau\in (-\pi /2,\pi /2),
                   \label{eq:2a}
\end{equation} 
with a real $\cT$-violation parameter $\tau$. Maximal $\cT$ violation occurs for $\tau =\pm\pi /4$. In the $\cT$-invariant case $H^A_{12}=0$~\cite{prl_2007} one has $\tau=0$. The transformation to this basis is achieved with $\hat O_0=e^{i\Phi_0\sigma_y}$ where, in terms of the components of $\vec h$, $\tan (2\Phi_0)=\Im\left(h_3/h_2\right)/\Im\left(h_1/h_2\right)$. The fitting procedure was tested thoroughly in~\cite{prl_2011}, via comparison with the method used in~\cite{prl_2001} to determine the complex eigenvalues of $H^{\rm eff}$. With Eq.~(\ref{eq:2}) the complex eigenvalues $E_j=f_j-i\Gamma_j/2$ of $H^{\rm eff}$ are given as
\begin{eqnarray}
E_{1,2}&=&\frac{f_1+f_2}{2}-i\frac{\Gamma_1+\Gamma_2}{4}\pm\sqrt{\cD}\nonumber\\
    \cD&=&\lvert\Re\vec h\rvert^2-\lvert\Im\vec h\rvert^2+2i\Re\vec h\cdot\Im\vec h\, .
                    \label{eq:4}
\end{eqnarray}       
At an EP the radicand $\cD$ vanishes, so that $\lvert\Re\vec h\rvert^2=\lvert\Im\vec h\rvert^2$ with $\lvert\Re\vec h\rvert^2,\, \lvert\Im\vec h\rvert^2$ nonvanishing, and $\Re\vec h\cdot\Im\vec h=0$. The EP was found by proceeding as described in~\cite{prl_2011}. It was confirmed by determining the geometric phases gathered by the eigenvectors of $H^{\rm eff}$ on encircling the EP.

The real parts of the eigenvalues yield the positions, the imaginary parts the widths of the resonances in the measured spectra $\vert S_{ab}(f)\vert^2$. Since the system is dissipative, the widths $\G_{1,2}>0$ are nonvanishing and, consequently, the eigenvalues are complex in the whole parameter plane. Yet, we will demonstrate in the following, that there exists a curve in the parameter plane along which the eigenvalues $\cE_{1,2}=E_{1,2}+i(\Gamma_1+\Gamma_2)/4$ of the non-Hermitian Hamiltonian $\cH=H^{\rm eff}+i\II (\Gamma_1+\Gamma_2)/4$ are either real or complex conjugate. 
This entails \cite{cmb-pt-prl-1998,cmb-rev,bender-berry-mandilara} that $\cH$ has a $\cP\cT$ symmetry along this curve, which is of so called {\em passive} type. The situation is similar to that in the experiments with optical wave-guides without active laser pumping \cite{guo-passive} and in Bose-Einstein condensates with leakage and without injection \cite{eva-pssive-pt-pra-2010}. A pair of real eigenvalues typical for $\cP\cT$-symmetric systems would correspond to resonances of zero width~\cite{most-prl-spec-sing-prl1}. It is obtained only after a suitably chosen width-offset. This can be avoided by a pumping mechanism~\cite{gain-loss5,Kottos_2011,schomerus-prl}. However, the eigenvectors and thus the physics of $\cP\cT$ symmetry and of the phase transition to spontaneously broken $\cP\cT$ symmetry at an EP is not affected by this shift. Furthermore, the dissipation and, consequently, the shift depends only marginally on the parameters $s$ and $\delta$, and thus the excitation frequency. Accordingly, we will consider $\cH$ instead of $H^{\rm eff}$. Note that $\hat T$ does not commute with the non-Hermitian $\cH$, not even when the internal Hamiltonian $H$ is $\cT$ invariant, as is the case for $B=0$. Yet there is a curve in the parameter plane on which $\cH$ is $\cP\cT$ symmetric. 

\begin{figure}[ht]
        \centering
        \includegraphics[width=7cm]{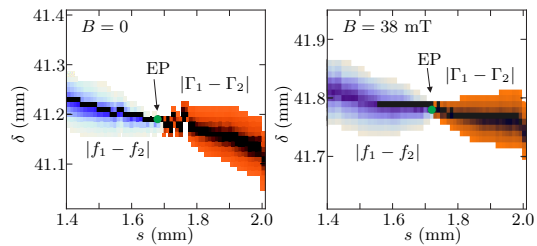}
        \caption{(Color online) Differences of the complex eigenvalues $E_{1,2}=f_{1,2}-i\Gamma_{1,2}/2$ in an area of the 
        parameter plane $(s,\delta)$ around the EP located at $(s_{\rm EP},\delta_{\rm EP})=(1.68\pm 0.01,41.19\pm 0.01)$~mm for $B=0$ (left panel) and at $(s_{\rm EP},\delta_{\rm EP})=(1.72\pm 0.01,41.78\pm 0.01)$~mm for $B=38$~mT (right panel).
        The darker the color the smaller is the respective difference.} 
        \label{fig:2}
\end{figure}

\begin{figure}[ht]
        \centering
        \includegraphics[width=8cm]{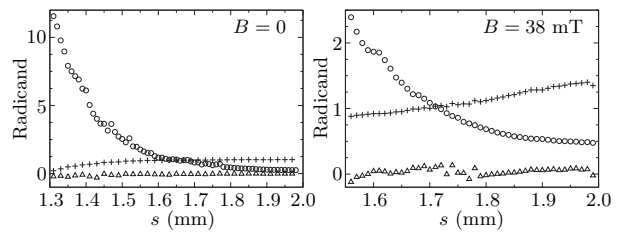}
        \caption{Radicand $\cD$ defined in Eq.~(\ref{eq:4}) for the parameter values $(s,\delta)$ along the curve of darkest color in Fig.~\ref{fig:2}. Shown are $\lvert\Re\vec h\rvert^2$ (crosses), $\lvert\Im\vec h\rvert^2$ (circles) and $\Re\vec h\cdot\Im\vec h$ (triangles), in the left panel for $B=0$, in the right one for $B=38$~mT. To obtain a dimensionless $\vec h$, it was divided by $\vert h_1\vert$.}
        \label{fig:3}
\end{figure}

{\em Experimental results.}\quad We located the curve in question by plotting for a given $B$ the difference $E_1-E_2=\cE_1-\cE_2=2\sqrt{\cD}$ in the parameter plane. The left and right panels of Fig.~\ref{fig:2} show the distances of the eigenvalues for 
$B=0$ and $B=38$~mT, respectively. The darker the color the smaller is the respective distance. The color scale was chosen such that only distances $\vert f_1-f_2\vert<3$~MHz and $\vert\Gamma_1-\Gamma_2\vert<0.35$~MHz are shown colored. We observe, that in both panels the distances of the real parts of the eigenvalues are small to the left side of the EP, i.e. for $s<s_{\rm EP}$, those of the imaginary parts for $s>s_{\rm EP}$ and both vanish at the EP, $(s,\delta)=(s_{\rm EP},\delta_{\rm EP})$.  Along the curve of darkest color they are vanishingly small. Figure~\ref{fig:3} shows the three parts of $\cD$ defined in Eq.~(\ref{eq:4}), $\lvert\Re\vec h\rvert^2,\, \lvert\Im\vec h\rvert^2$ and $\Re\vec h\cdot\Im\vec h$, separately along this curve,
in the left panel for $B=0$ and in the right one for $B=38$~mT. Note the continuity of the curves, which demonstrates the precision of the measurements and of the determination of $H^{\rm eff}$. In fact, each point was obtained from an independent fitting. For both values of $B$, $\Re\vec h\cdot\Im\vec h$ (triangles) is vanishingly small, that is the radicand $\cD$ is approximately real along the dark curves in Fig.~\ref{fig:2}. The curves $\lvert\Re\vec h\rvert^2$ (crosses) and $\lvert\Im\vec h\rvert^2$ (circles) cross at the EP. For $s>s_{\rm EP}$ the former is larger than the latter; consequently $\sqrt{\cD}$ and thus the eigenvalues of $\cH$ are real. They change into a pair of complex conjugate eigenvalues for $s<s_{\rm EP}$. The transition takes place at the EP. 
This behavior is reminiscent of that of a $\cP\cT$-symmetric Hamiltonian complemented by a spontaneous breaking of the $\cP\cT$ symmetry of its eigenvectors at the EP.

{\em Interpretation of the experiment.}\quad To reveal the $\cP\cT$ symmetry of $\cH$ along the curve $\Re\vec h\cdot\Im\vec h =0$ we search for the basis transformation which yields the parity operator as $\hat P=\sigma_x$. Most generally, in this basis a Hamiltonian that commutes with $\hat P\hat T$ can be written in the form
\begin{equation}
H^{\rm PT}=\left(\begin{array}{cc}
         {\rm A}+i{\rm B}\,          &{\rm C}+i{\rm D}\\
         {\rm C}-i{\rm D}\,          &{\rm A}-i{\rm B}
        \end{array}
  \right)\, {\rm with}
{\rm \, A,\, B,\, C,\, D}\, \in\RR\, .
                   \label{eq:5}
\end{equation}
The eigenvalues of $H^{\rm PT}$ are real when its eigenvectors also commute with $\hat P\hat T$, complex conjugate otherwise.
If $\cT$ invariance is violated $H_{12}^A$ in Eq.~(\ref{eq:2}) and $\tau$ in Eq.~(\ref{eq:2a}) are nonvanishing. The $\cT$-violation parameter $\tau$ varies with the parameters $s$ and $\delta$, because the wave functions~\cite{prl_2001,Stock_2010} and thus the electromagnetic field at the position of the ferrite change with the opening size $s$ of the slit and the position $\delta$ of the Teflon piece. In Fig.~{\ref{fig:4} we show for $B=38$~mT its change along the curve characterized by $\Re\vec h\cdot\Im\vec h =0$. 
\begin{figure}[!t]
        \centering
        \includegraphics[width=4.5cm]{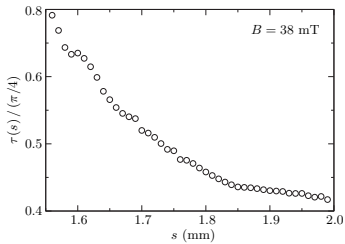}
        \caption{The parameter $\tau$ in units of maximum $\cT$ violation $\pi/4$ along the curve of darkest color in Fig.~\ref{fig:2} for $B=38$~mT.}
        \label{fig:4}
\end{figure}
On this curve $\cH$ can be brought into the form of Eq.~(\ref{eq:5}). The transformation $\hat U=e^{-i(\tau/2)\sigma_z}=\left(\begin{array}{cc}
         \exp(-i\tau/2)\,          &0\\
         0\,          &\exp(i\tau/2)
        \end{array} \right)$ turns $\cH$ into a complex symmetric matrix. For a $\cT$-invariant scattering process $\tau=0$ and $\hat U=\II$, because then $\cH$ is already complex symmetric [see  Eq.~(\ref{eq:2})]. 
If and only if $\cD$ is real, i.e. $\Re\vec h\cdot\Im\vec h =0$, there is a basis transformation that brings $\hat U\cH\hat U^\dd$ to the form of Eq.~(\ref{eq:5}). This is achieved with the real orthogonal transformation $\hat O=e^{i\Phi\sigma_y}$, where $\tan(2\Phi)=\Im h_1/\left[\Im h_3\cos\tau\right]$. The transformation yields Eq.~(\ref{eq:5}) with ${\rm A} =(f_1+f_2)/2$, ${\rm B} =\Im h_1/\left[\sin\left(2\Phi\right)\cos\tau\right]$, ${\rm C} =\Re h_1/\left[\cos\left(2\Phi\right)\cos\tau\right]$ and ${\rm D}=0$. 

{\em Conclusions.}\quad We experimentally identified a curve in the parameter plane along which the eigenvalues of $\cH$ are either real or complex conjugate. There, the radicand $\cD$ is real, i.e. $\Re\vec h\cdot\Im\vec h =0$. This relation is independent of the basis representation of $\cH$. We have specified the basis transformation, which brings $\cH$ to the form of Eq.~(\ref{eq:5}) along the curve. In that basis, $\cP\cT$ symmetry corresponds to invariance with respect to $\hat P\hat T$. If the scattering system is $\cT$ invariant, that is for $B=0$, the basis change is achieved with the real orthogonal transformation $\hat O$, i.e.~$[\hat P\hat T,\hat O\cH\hat O^T]=0$. In distinction to previous experiments~\cite{gain-loss5,guo-passive,Kottos_2011}, the present data include $\cT$ violation of the scattering process, induced with an external magnetic field $B\ne 0$. In that case, the unitary basis transformation $\hat O\hat U$ brings $\cH$ to the form of Eq.~(\ref{eq:5}), that is, $[\hat P\hat T,\hat O\hat U\cH\hat U^\dd\hat O^T]=0$. Thus, $\cH$ is invariant under the antilinear operator $\hat U^\prime\hat T$, where $\hat U^\prime=\hat U^\dd\hat O^T\hat P\hat O\hat U^\dd$ is unitary, in accordance with~\cite{bender-berry-mandilara}. As predicted, the change from real eigenvalues for $s>s^{EP}$ to complex conjugate ones for $s<s^{EP}$ is accompanied by a sponteneous breaking of $\cP\cT$ symmetry of the eigenvectors of $\hat O\hat U\cH\hat U^\dd \hat O^T$ at the EP, that is, they cease to be eigenvectors of $\hat P\hat T$~\cite{cmb-pt-prl-1998,cmb-rev}. 

Illuminating discussions with M. V. Berry and H. A. Weidenm\"uller are gratefully acknowledged. This work was supported by the DFG within SFB 634.



\begin{thebibliography}{24}
\bibitem{QB}
H.-J. St\"ockmann and J. Stein, Phys. Rev. Lett. \textbf{64}, 2215 (1990);
H.-D. Gr\"af {\it et~al.}, Phys. Rev. Lett. \textbf{69}, 1296 (1992).

\bibitem{Richter}
S. Sridhar, Phys. Rev. Lett. {\bf 67}, 785 (1991);
A. Richter, in \textit{Emerging Applications of Number Theory}, edited by
D. A. Hejhal {\it et~al.}, IMA Vol. 109 (Springer, NY, 1999), p. 479;
H.-J. St{\"o}ckmann, \emph{Quantum Chaos: An Introduction}
(Cambridge University Press, Cambridge, 2000).

\bibitem{cmb-pt-prl-1998} C. M. Bender and S. Boettcher, Phys. Rev. Lett. {\bf 80}, 5243 (1998).
\bibitem{bender-berry-mandilara}C. M. Bender, M. V. Berry, and A. Mandilara, J. Phys. A {\bf 35}, L467 (2002).
\bibitem{cmb-rev} C. M. Bender, Rep. Prog. Phys. {\bf 70}, 947 (2007).
\bibitem{kato} T. Kato: {\it Perturbation theory for linear operators} (Springer, Berlin,
1966).

\bibitem{PT-brach} C. M. Bender, D. C. Brody, H. F. Jones, and B. K. Meister, Phys. Rev. Lett. {\bf 98}, 040403 (2007); U. G\"unther and B. F. Samsonov, Phys. Rev. Lett. {\bf 101}, 230404 (2008).

\bibitem{PT_Theorie}R. El-Ganainy, K. G. Makris, D. N. Christodoulides, and Z. H. Musslimani, Opt. Lett. {\bf 32},  2632 (2007); K. G. Makris, R. El-Ganainy, D. N. Christodoulides, and Ziad H. Musslimani,Phys. Rev. Lett. {\bf 100}, 103904 (2008); Phys. Rev. A {\bf 81}, 063807 (2010); Z. H. Musslimani, K. G. Makris, R. El-Ganainy, and D. N. Christodoulides, Phys. Rev. Lett. {\bf 100}, 030402 (2008).

\bibitem{gain-loss5} C. E. R\"uter \etal,  Nature Physics {\bf 6}, 192 (2010).
\bibitem{guo-passive}A. Guo \etal, Phys. Rev. Lett. {\bf 103}, 093902 (2009).

\bibitem{gain-loss3} S. Klaiman, U. G\"unther, and N. Moiseyev,  Phys. Rev. Lett. {\bf 101}, 080402 (2008);
E.-M. Graefe and H. F. Jones, Phys. Rev. A {\bf 84}, 013818 (2011).

\bibitem{Kottos_2011} J. Schindler, A. Li, M. Zhang, F. M. Ellis, and T. Kottos, Phys. Rev. A {\bf 84}, 040101 (2011).

\bibitem{most-prl-spec-sing-prl1}A. Mostafazadeh, Phys. Rev. Lett. {\bf 102}, 220402 (2009).
\bibitem{most-prl-spec-sing-pra2}A. Mostafazadeh, Phys. Rev. A {\bf 83}, 045801 (2011).

\bibitem{stone-laser-absorber-prl}Y. D. Chong, L. Ge, and A. D. Stone, Phys. Rev. Lett. {\bf 106}, 093902 (2011).

\bibitem{dim-kott} H. Ramezani, T. Kottos, R. El-Ganainy, and D. N. Christodoulides, Phys. Rev. A {\bf 82}, 043803 (2010);
Z. Lin \etal, Phys. Rev. Lett. {\bf 106} 213901 (2011).

\bibitem{longhi-bloch-osc-prl}S. Longhi, Phys. Rev. Lett. {\bf 103}, 123601 (2009).
\bibitem{segev-pt-graphene-tach}A. Szameit, M. C. Rechtsman, O. Bahat-Treidel, and M. Segev, Phys. Rev. A {\bf 84}, 021806 (2011).

\bibitem{schomerus-prl}H. Schomerus, Phys. Rev. Lett. {\bf 104}, 233601 (2010); Phys. Rev. A {\bf 83}, 030101 (2011).

\bibitem{prl_2001}
C. Dembowski {\it et~al.}, Phys. Rev. Lett. \textbf{86}, 787 (2001).

\bibitem{prl_2011} B.~Dietz {\it et~al.}, Phys.\ Rev.\ Lett.\ {\bf 106}, 150403 (2011).

\bibitem{prl_2007} B.~Dietz {\it et~al.}, Phys.\ Rev.\ Lett.\ {\bf 98}, 074103 (2007).

\bibitem{mahaux-weidenmueller} C. Mahaux and H. A. Weidenm\"uller, {\it Shell-model approach to nuclear reactions}, (North-Holland Publ. Comp., Amsterdam, 1969).

\bibitem{prl_2009} B. Dietz {\it et~al.}, Phys. Rev. Lett. {\bf 103}, 064101 (2009). 

\bibitem{Albeverio} S. Albeverio {\it et~al.}, J. Math. Phys. {\bf 37}, 4888 (1996).

\bibitem{prl_stoeckmann03} R. A. M{\'e}ndez-S{\'a}nchez {\it et~al.}, Phys. Rev. Lett. {\bf 91}, 174102 (2003).

\bibitem{henley} H. Frauenfelder and E. M. Henley, {\it Nuclear and Particle Physics} (W.A. Benjamin, Reading, MA, 1975); F. Coester, Phys. Rev. {\bf 84}, 1259 (1951).

\bibitem{eva-pssive-pt-pra-2010}E.-M. Graefe, H. J. Korsch, and A. E. Niederle, Phys. Rev. A {\bf 82}, 013629 (2010).
\bibitem{Hernandez} E. Hern\'andez and A. Mondrag\'on, Phys. Lett. B {\bf 326}, 1 (1994).
 
\bibitem{Stock_2010} B. K{\"o}ber {\it el~al.}, Phys. Rev. E {\bf 82}, 036207 (2010).
\end{thebibliography}
\end{document}